\documentclass[final]{iopart}

\usepackage{hyperref}
\usepackage{iopams}
\usepackage[dvips]{graphicx}
\usepackage{color}

\begin{document}

\title{Broadband optical isolator in fibre optics}

\author{Micha\l $\,$ Berent} 
\address{Faculty of Physics, Adam Mickiewicz University, Umultowska 85, 61-614 Pozna\'{n}, Poland}
\ead{mkberent@gmail.com}

\author{Andon A. Rangelov and Nikolay V. Vitanov}
\address{Department of Physics, Sofia University, James Bourchier 5 Blvd., 1164 Sofia, Bulgaria}

\begin{abstract}
We propose a broadband optical diode, which is composed of one
achromatic reciprocal quarter-wave plate and one non-reciprocal quarter-wave
plate, both placed between two crossed polarizers. The presented design of
achromatic wave plates relies on an adiabatic evolution of the Stokes
vector, thus, the scheme is robust and efficient. The possible simple implementation using fibre
optics is suggested.
\end{abstract}



\submitto{Journal of Optics}

%with maketitle only a title is on the first page
%\maketitle

%======================

\section{Introduction}
\label{intro}

%======================

An optical isolator (optical diode) is an optical component that allow the
light to pass in one direction but block it in the opposite direction. These
devices are commonly used in laser technology to prevent the unwanted
backreflections which might be harmful to optical instrumentation. The
standard optical isolator, as first proposed by Rayleigh \cite{Rayleigh1885}%
, is composed of two polarizers with their transmission axes rotated by $%
45^{\circ}$ with respect to each other and a Faraday rotator. The Faraday rotator is
made of a magnetoactive medium which is placed inside a strong magnet. The
magnetic field induces a circular anisotropy in the material (Faraday
effect), which makes the left and right circular polarizations experience a
different refraction index. As a result, the plane of linear polarization
travelling through the device is rotated by an angle equal to
\begin{equation}
\theta(\lambda) = \nu(\lambda) B L,  \label{rotation angle}
\end{equation}
where $B$ is the induction of the applied magnetic field, $L$ is a length of
the magnetoactive medium and $\nu(\lambda)$ is the Verdet material constant.
Because the Verdet constant depends strongly on the wavelength, so does the
rotation angle of the rotator.

\begin{figure}[h]
\centering
\includegraphics[width=.5\columnwidth]{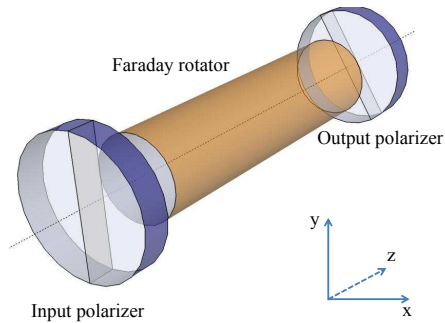}
\caption{(Colour online) The scheme of a standard Faraday isolator. The light travelling forwards is first polarized in the vertical direction, then the plane of polarization is rotated by $45^{\circ}$ by the Faraday rotator and passes through the output polarizer. The light travelling backwards is first polarized at $45^{\circ}$, then it is rotated by another $45^{\circ}$ in the Faraday element. The resulting horizontal polarization is extinguished by the input vertical polarizer.}
\label{faraday}
\end{figure}

The standard optical diode shown in \Fref{faraday} works as follows. The
light travelling in the forward direction is first linearly polarized in the
horizontal direction by the input polarizer. Afterwards, the Faraday element
rotates the polarization by $45^{\circ}$ and finally, the light is
transmitted through the output polarizer. The light travelling backwards is
first linearly polarized at $45^{\circ}$, the Faraday rotator then rotates
the polarization by another $45^{\circ}$, meaning the light is now polarized
in the vertical direction. Because the input polarizer transmits only
horizontal polarization, the light is extinguished. The main drawback of
standard isolators is that they work efficiently only for a very narrow
range of wavelengths, because of the dispersion of the Faraday rotation
angle $\theta(\lambda)$.

%===============================================================================

The usual approach seen in commercial broadband isolators is to add the
additional reciprocal rotator (e.g. quartz rotator) next to the Faraday rotator.
The former element is used to compensate for the dispersion of the Faraday
rotator \cite{Iwamura1978, Proffitt1980, Schulz1989, Schulz1991,
Parfenov2002}. In the forward direction, the rotation of the two elements add
up to a total rotation of $90^{\circ}$, whereas in the opposite direction
the rotations subtract and no rotation is experienced by the plane of linear
polarization.

Recently we proposed a novel broadband isolator \cite{Berent2013}, which could
be realized with the bulk optics elements. We exploited the analogy in the
mathematical description of a quantum two-state system driven by a pulsed
laser field and an electromagnetic wave propagating through an anisotropic
medium. The technique of composite pulses known from nuclear magnetic
resonance (NMR) \cite{Levitt1979, Levitt1986} and quantum
optics \cite{Torosov2011} was applied by us to find conditions for
broadband operation of the isolator.

In this article, we propose an alternative realization of the optical isolator
which could be suitably implemented in fibre optics. Our approach is based
on the adiabatic evolution of the Stokes vector \cite{Rangelov10, Kundikova1997} which allows for a broadband performance of the
presented isolator.

As high-power fibre lasers are attracting an increasing attention, the
development of integrated optical elements for the manipulation of the state
of light becomes a crucial point. An all-fibre architecture has the
advantage of allowing for an efficient transmission of light without
reflections losses on the way (except from an input to a fibre); such losses are a
serious problem in bulk optics. As the broadband high-power sources like
superluminescent diodes (SLD) or Ti:Sapphire oscillators are broadly used in
optical coherence tomography (OCT) \cite{Fujimoto1993, Fujimoto1995},
characterization of optical components \cite{Fujimoto1992} and optical
measurements \cite{Hansch2002}, the issue of the efficient broadband
isolation in fibres is of increasing importance.

The composition of the manuscript is the following. In section \ref{stokes}
we present the mathematical description underpinning our approach. Section \ref{design} discusses the design of our broadband optical isolator.
Then we consider the practical realization of the proposed design in section %
\ref{practical}. Section \ref{results} presents the performance of the
diode and in the last section \ref{conclusion} we summarize the conclusions.

%=============================================

\section{Stokes formalism}
\label{stokes}
%=============================================

Consider the propagation of a plane electromagnetic wave
through an anisotropic dielectric medium along the $z$-axis. We assume that there are no
polarization dependent losses. Then the equation of motion is given by the torque equation \cite%
{MacMaster1961, Schmieder1969, Kubo1980, Kubo1983, Kubo1985}
\begin{equation}
\frac{d}{dz}\mathbf{S}(z)=\mathbf{\Omega }(z)\times \mathbf{S}(z)
\label{torque equation}
\end{equation}%
where $\mathbf{S}(z)=\left[ S_{1}(z),S_{2}(z),S_{3}(z)\right] $ is the Stokes
polarization vector representing any state of polarization on the Poincar\'{e}
sphere, and $\mathbf{\Omega }(z)=\left[ \Omega _{1}(z),\Omega _{2}(z),\Omega
_{3}(z)\right] $ is a birefringence vector of the medium. One can write down \Eref{torque equation} in matrix form as%
\begin{equation}
\frac{d}{dz}\mathbf{S}(z)=\mathbf{H}(z)\cdot\mathbf{S}(z),
\end{equation}%
where the matrix $\mathbf{H}(z)$ is given as
\begin{equation}
\mathbf{H}(z)=\left[
\begin{array}{ccc}
0 & -\Omega _{3}(z) & \Omega _{2}(z) \\
\Omega _{3}(z) & 0 & -\Omega _{1}(z) \\
-\Omega _{2}(z) & \Omega _{1}(z) & 0%
\end{array}%
\right] \,.
\end{equation}%
We shall make use of the adiabatic evolution of the Stokes vector. For this purpose,
we need the eigenvalues of $\mathbf{H}(z)$, which read%
\begin{equation}
\varepsilon _{-}(z)=-i\left\vert \Omega (z)\right\vert ,\ \ \varepsilon
_{0}(z)=0,\ \ \varepsilon _{+}(z)=i\left\vert \Omega (z)\right\vert ,
\label{eigenvalues}
\end{equation}%
with $\left\vert \Omega (z)\right\vert =\sqrt{\Omega _{1}^{2}(z)+\Omega
_{3}^{2}(z)+\Omega _{3}^{2}(z)}$. The eigenvector that corresponds to the
zero eigenvalue is extremely simple:
\begin{equation}
\sigma (z)=\frac{\Omega _{1}(z)S_{1}(z)+\Omega _{2}(z)S_{2}(z)+\Omega
_{3}(z)S_{3}(z)}{\left\vert \Omega (z)\right\vert }.
\label{Stokes dark state}
\end{equation}%
We will call this eigenvector ``polarization dark state" in analogy to the
stimulated Raman adiabatic passage (STIRAP) process in quantum optics \cite%
{Gaubatz1990,Bergmann1998,Bergmann2001}. Assuming that the evolution is adiabatic
and that the Stokes polarization vector $\mathbf{S}(z)$ is initially aligned with
the polarization ``dark state" $\sigma (z)$, then the Stokes vector will
follow this adiabatic state throughout the medium. The evolution of the polarization ``dark
state" depends on the initial polarization and the spatial ordering of
the components of birefringence vector. It will be discussed in detail in
the next section.

In analogy to the quantum-optical STIRAP \cite{Bergmann1998,Bergmann2001} the
condition for adiabatic evolution requires  the integral of the length of birefringence vector over the
propagation distance $L$ to be large, i.e.
\begin{equation}
\int_{0}^{L}\left\vert \Omega (z)\right\vert dz\gg 1\,.  \label{adiabatic}
\end{equation}%

%===============================================
\section{The design of wave-plates}

\label{design}
%================================================

The optical isolator we are going to present requires two crossed
polarizers and two achromatic optical elements: a
reciprocal (standard) quarter wave plate and a non-reciprocal quarter wave
plate. Below we will describe the design in the framework of formalism presented above.

%
%===============================================================================
\begin{figure}[h]
\centerline{\includegraphics[width=.5\columnwidth]{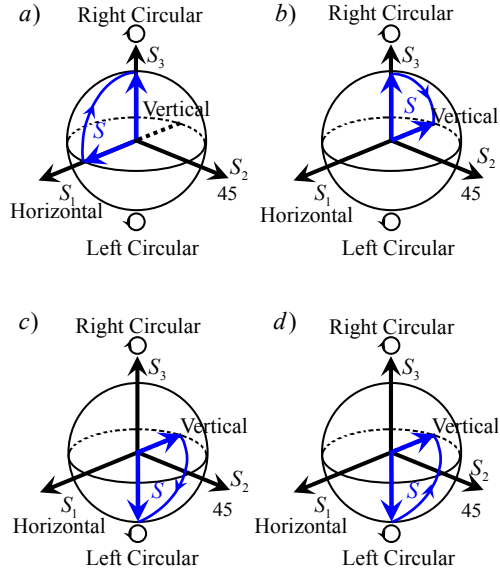}}
\caption{(Colour online) Non-reciprocal polarization transformation propagation. Top frames
demonstrate the forward polarization evolution along the direction of the
fixed magnetic field: a) Starting from horizontally polarized light and passing through
the achromatic reciprocal wave plate leads to right-circular polarization. b)
Polarization is evolved from right-circular polarization to vertical, due to the achromatic non-reciprocal wave plate. Bottom frames
demonstrate the backward polarization evolution against the direction of the
fixed magnetic field: c) Starting from vertical polarization and passing
through the achromatic non-reciprocal wave plate the light evolves into
left-circular polarization. d) Light pass through the achromatic reciprocal wave
plate and, thus, it is returned to vertical polarization.}
\label{evolution}
\end{figure}

\subsection{Reciprocal quarter wave-plate}

\label{reciprocal}

Let us first analyze the operation of the achromatic reciprocal quarter
wave-plate. This problem was recently studied in \cite{Huang1997a,
Huang1997b}. The reciprocity of the wave plate comes from the reciprocity of
the birefringence vector. This means that when the light travels
through the wave plate in the reverse direction, the sign of the
birefringence vector is also reversed.

Bearing this in mind, we analyze the evolution of the polarization
 dark state \Eref{Stokes dark state}. Let us assume that
initially the light is linearly polarized in the horizontal direction, $\mathbf{S}(z_{i})=\left[ 1,0,0\right] $. When $\Omega _{1}(z)$ precedes $%
\Omega _{2}(z)$ and $\Omega _{3}(z)$, and the Stokes vector smoothly
follows the birefringence vector, the polarization ends up in the right
circular polarization state $\mathbf{S}(z_{f})=\left[ 0,0,1\right] $, provided that only $\Omega _{3}(z)>0$ is present at the end (e.g. through a large spin rate of the fibre \cite{Huang1997a, Huang1997b}). The process is fully
reversible, meaning that if we change the ordering of the birefringence vector
components ($\Omega _{3}(z)$ now precedes $\Omega _{1}(z)$ and $\Omega_{2}(z)$), and the Stokes vector is initially aligned along $S_{3}$-axis, it
adiabatically evolves into state $\mathbf{S}(z_{f})=\left[ 1,0,0\right] $
(horizontal linear polarization). The latter holds if only $\Omega _{1}(z_f)>0$ is
present at the end. Thus, this arrangement operates as the standard quarter wave plate.

\subsection{Non-reciprocal quarter wave-plate}
\label{non-reciprocal}

The second case of the non-reciprocal quarter wave plate requires the use of a
non-reciprocal birefringent element, which would be sensitive to the
direction of propagation of light. It is long known that the magnetic field
applied to a magnetoactive medium induces a circular birefringence which
makes the left and right circularly polarized light experience different
refractive indices resulting in rotation of the plane of linear polarization.

In the forward direction the operation of the element with one of the
birefringence components being non-reciprocal is the same as for a
reciprocal one: the light initially polarized horizontally $\mathbf{%
S}(z_i) = \left[1,0,0\right]$ is transformed into the right circular
polarization state $\mathbf{S}(z_f) = \left[0,0,1\right]$. However, the
propagation in the reverse direction is different. If we start with the
right circular polarization $\mathbf{S}(z_i) = \left[0,0,1\right]$ it
evolves into the linear \textit{vertical} polarization $\mathbf{S}(z_f) = %
\left[-1,0,0\right]$ (see \Fref{evolution}).

%==================================
\section{Practical implementation}
\label{practical}
%==================================

The design of the broadband optical isolator described in section \ref%
{design} could be conveniently implemented with single-mode optical fibre.
The use of fibre-optic isolator is exceedingly attractive for
integrated fibre-optic systems as well as high power applications.

\begin{figure}[h]
\centerline{\includegraphics[width=.5\columnwidth]{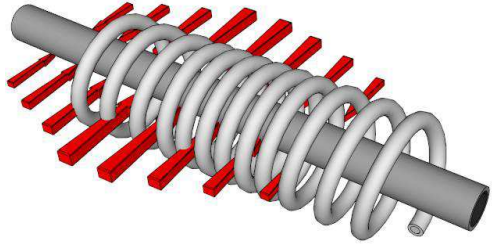}}
\caption{(Colour online) Fibre optics setup of the achromatic quarter wave plate with overlapping linear stress-induced
birefringence and circular birefringence generated through the Faraday
effect (magnetic field along the fibre). The magnitude of
the horizontal stress is represented by red arrows varying in length.
Analogously, the number of turns in the coil indicate pulse-shaped spatial
variations of the longitudinal magnetic field.}
\label{fiber}
\end{figure}

The single-mode fibre has
to exhibit both linear and circular birefringence. The first one might be
induced by stress applied to a fibre or by external electric field \cite%
{Kundikova1997, Fernandes2012}. To induce the circular birefringence one
might apply a torsion of the fibre (for the reciprocal effect) or the external magnetic field (through
the Faraday non-reciprocal effect). The possible implementation is depicted in \Fref{fiber}.

The achromatic wave-plate described in section \ref{reciprocal} would be implemented in the single-mode fibre with the combined stress-induced
linear birefringence and torsion of the fibre which would generate a
circular birefringence. The choice of torsion-induced circular birefringence
is dictated by the requirement that this part of our setup must be
reversible.

The first author that experimentally demonstrated the achromatic
and adiabatic quarter-wave plate was Huang \cite{Huang1997a,
Huang1997b}. He used a spun fibre with the spinning rate increased
with the distance. Those reciprocal designs allowed for the
transformation of polarization from linear to circular and back.
The reciprocal achromatic quarter-wave plate could be
alternatively realized with commercially available achromatic
quarter-wave plate.

The non-reciprocal achromatic quarter wave plate can be made similarly to
the reciprocal one. The only difference is that one has to use circular
birefringence which would be non-reciprocal with respect to the direction of
propagation of light. As pointed out earlier, the magnetic field generates
such birefringence through the Faraday effect. Similarly to the Faraday
rotator in the standard isolator, this element is crucial for the design
of the practical optical diode. The problem with the Faraday effect in
standard silica fibres is that the Verdet constant of the medium is very low
\cite{Cruz1996} and, thus, one needs a very long piece of fibre or very
strong magnetic field to achieve a significant rotation. Many different
approaches has been devised to overcome this difficulty. It seems that the
most successful is the doping of fibre with rare-earth ions like terbium ($
Tb^{3+}$). The value of Verdet constant achievable with these fibres \cite{Ballato1995, Sun2010t, Sun2010} is nearly as high as that of the bulk
optics rotators made of Terbium Gallium Garnet (TGG) \cite{Grumman}.

%===============================================================================

%===============================================================================
\section{Results}
\label{results}

%===============================================================================
\begin{figure}[h]
\centerline{\includegraphics[width=.5\columnwidth]{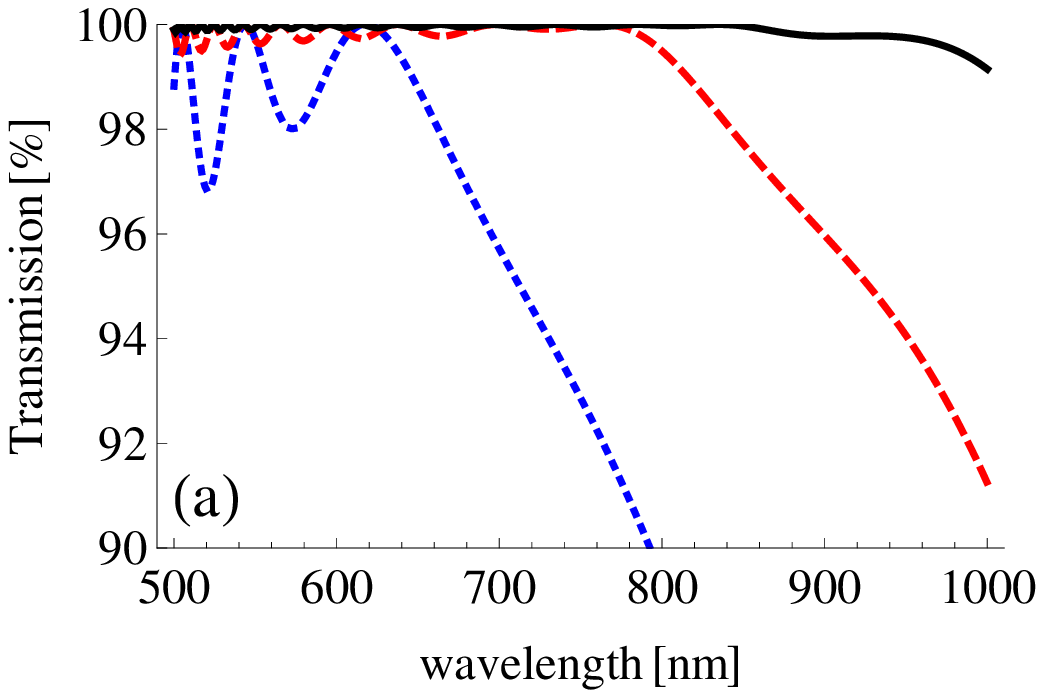}} %
\centerline{\includegraphics[width=.5\columnwidth]{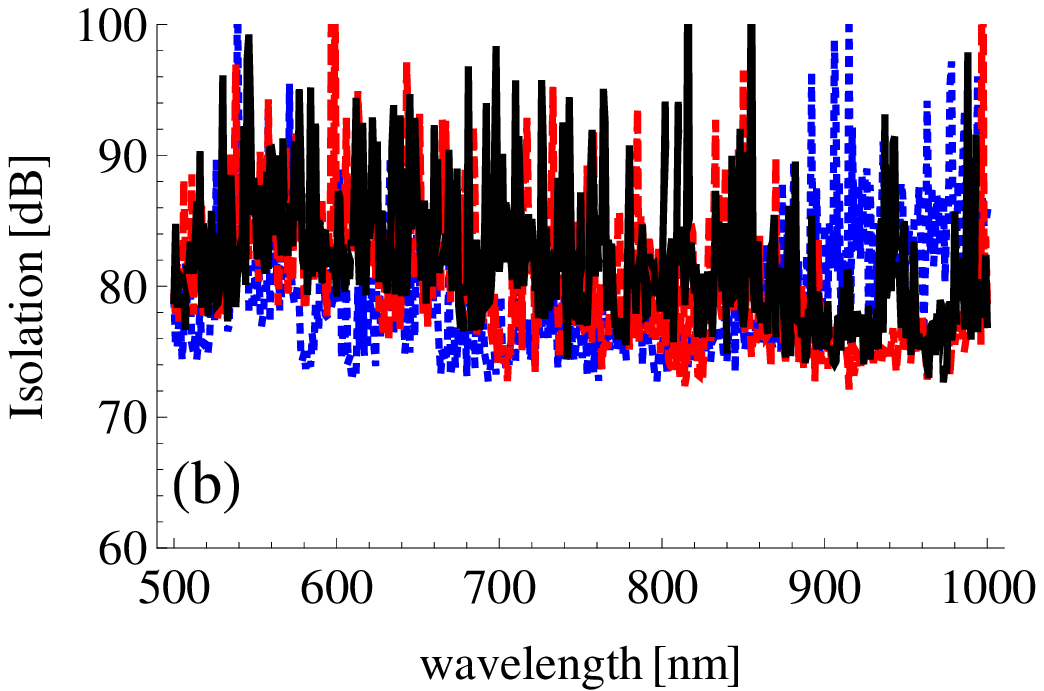}} %
%\centerline{\includegraphics[width=.5\columnwidth]{figure4c.eps}}
\caption{(Colour online) Simulation of the performance of the broadband fibre optical
isolator for three different lengths $L_1 = 1$~m (black solid line), $%
L_2 = 0{.}5$~m (red dashed line) and $L_3 = 0{.}2$~m (blue dotted line). We
plotted (a) the transmission of the isolator (intensity transmitted in
the forward direction), and (b) the isolation of the optical diode, versus
wavelength.}
\label{isotrans}
\end{figure}

We performed numerical simulations of the performance of the design we
described in this manuscript. In our calculation we assumed the
stress-induced birefringence equal to $\beta = \Delta n = 10^{-5}$ which is
easily achievable with the existing technology \cite{Fernandes2012}. The
linear birefringence component of the birefringence vector
is then given by the equation
\begin{equation}
\Omega_1(z) = \frac{2 \pi \beta}{\lambda} \cos\left(\frac{\pi z}{2 L}\right),
\end{equation}
where $L$ is the length of the proposed in-fibre isolator.

As mentioned before, the Verdet constant of the standard silica fibre is
very low. Thus, we decided to consider the silica fibre doped with
paramagnetic terbium ions. The dispersion of such fibre is similar to a bulk
optics TGG crystal \cite{Sawanobori2002, Villora2011}
\begin{equation}
\nu(\lambda) = \frac{A}{\lambda^2-\lambda_0^2},
\end{equation}
with slightly different fit parameters $A = -19{.}7\cdot 10^6\,$(nm$^2$~rad)/(T$\cdot$m) and $\lambda_0 = 385$~nm \cite{Ballato1995}. The
circular component of the birefringence vector now reads
\begin{equation}
\Omega_3(z) = \nu(\lambda) B_0 \sin\left(\frac{\pi z}{2 L}\right),
\end{equation}
with $B_0$ being the amplitude of the magnetic field induction. We assumed the magnitude of the
magnetic field to be $B_0 = 1$~T. Simulations were carried out for three different
lengths: $L_1 = 1$~m, $L_2 = 0{.}5$~m and $L_3 = 0{.}2$~m.

%\textcolor{red}{
The choice of optically active medium (TGG) roughly defines the wavelength range of presented optical isolator. Transmission window of terbium doped fibre (500-1000~nm) and the magnitude of Faraday rotation constitutes the range of operation of isolator simulated in this section. Such isolator would be suitable in experiments with Ti:Sapphire laser (central wavelength at 780~nm). To achieve a similar performance in a different wavelength range requires a different choice of paramagnetic ions, which would induce a Faraday rotation of significant amplitude. For a telecommunication wavelengths bismuth rare-earth ions might be an appropriate choice.
%}

%\textcolor{red}{
The other issue that must be taken into account is a single-mode performance of fibre. The higher order modes induced in a fibre might negatively affect the performance of isolator. Such fibre must be adjusted to a wavelength window in which the isolator is expected to operate. The commercial manufacturers provide a wide range of such fibres \cite{SMF}.
%}

We quantify the performance of the designed isolator with its transmission
and isolation. Transmission indicates the intensity of light passing through
the isolator as compared to the intensity at the input. In the forward
direction we have
\begin{equation}
T = \frac{I_{forw}}{I_0} = \frac{1}{2}\left(1+\mathbf{S}_{forw}^{T}(z_f)\mathbf{S}_{out}\right),,
\end{equation}
in the backward direction
\begin{equation}
B = \frac{I_{back}}{I_0} = \frac{1}{2}\left(1+\mathbf{S}_{back}^{T}(z_f)\mathbf{S}_{in}\right),
\end{equation}
where $\mathbf{S}_{forw}(z_f)$ represent the Stokes vector of light
travelling forwards,  $\mathbf{S}_{back}(z_f)$ --- for light travelling backwards, $\mathbf{S}_{in} = [1,0,0]$ and $\mathbf{S}_{out} = [-1,0,0]$ refer to the input horizontal and output vertical polarizers, respectively. Furthermore, $I_0$ is the intensity of light
entering the isolator, whereas $I_{forw}$ and $I_{back}$ are the intensities
measured after the diode in the forward and backward directions.

The isolation was calculated using the standard formula \cite{Berent2013, Adams2012}
\begin{equation}
D = -10\log\left(\frac{I_{back}}{I_{forw}}\right).
\end{equation}

\Fref{isotrans} depicts the results of our calculations for the three fibre lengths. In \Fref{isotrans}~a we presented the intensity of light in the forward direction and in \fref{isotrans}~b the isolation. One can notice the exceptionally high level of isolation over the whole range of wavelengths considered. What is interesting, the level of isolation is almost constant irrespective of the length of the isolator.

The price we have to pay for broadband isolation is the transmission window decreasing with decreasing length of the isolator. As the length of the setup decreases the adiabatic condition is weakened and, thus, the transmission becomes worse.

The transmission was calculated with an assumption that the fibre is lossless. In practice, the light travelling through a fibre is attenuated. However, because the length of our isolator is relatively short, the losses should be negligible.
The isolation of the best commercial broadband fibre diodes \cite{thorlabs}
is no greater than 32~dB and the range of isolation is around 150~nm. As
seen in \Fref{isotrans}~b, the isolation of our diode remains greater
than 70~dB for a range as wide as 500~nm. Because of the robustness of
adiabatic techniques, the isolation would be also insensitive to variations
in the temperature and the length of the fibre.

%======================

\section{Conclusions}
\label{conclusion}
%======================
In this manuscript we proposed a novel design of the fibre optical isolator,
which operates over a broad range of wavelengths. The adiabatic evolution
has been successfully applied to obtain a robust broadband performance of
the optical diode under study. The isolator can be further enhanced by
inducing birefringence of higher magnitude. This is possible with
stress-induced birefringence, as in our simulations we used a moderate value
of the former. To obtain higher circular birefringence with the Faraday
effect one can apply stronger magnetic field or use a different
magnetoactive medium to assure higher value of the Verdet constant.
Increasing the value of birefringence would also result in the decrease
of the length of the device.

%===============================
\section*{Acknowledgements}
This work is supported by the Bulgarian NSF Grant DMU-03/103.

\end{document}